\documentclass[11pt]{article}
\usepackage[margin=1in]{geometry}
\usepackage{amsmath,amssymb,bm}
\usepackage{graphicx}
\usepackage{booktabs}
\usepackage{array}
\usepackage{caption}
\usepackage{subcaption}
\usepackage{float}
\usepackage{microtype}
\usepackage{hyperref}
\usepackage{xurl}
\usepackage{xcolor}
\usepackage{siunitx}
\hypersetup{colorlinks=true,linkcolor=blue!55!black,citecolor=blue!55!black,urlcolor=blue!55!black}
\newcommand{\Lc}{\Lambda_c^+}
\newcommand{\Lb}{\bar\Lambda}
\newcommand{\Bbar}{\bar B^0}
\newcommand{\MeV}{\mathrm{MeV}}

\title{A Search in LHCb Open Data for $f$-Baryon Resonances Predicted at HADRON 2025}
\author{Scott Chapman\\[2pt]
\small Institute for Quantum Studies, Chapman University, Orange, CA 92866, USA}
\date{September 2026}

\begin{document}
\maketitle

\begin{abstract}
A prediction presented at HADRON 2025 proposed two baryon-like states containing an additional hypothesized $f$ quark that could appear in the decay $\Bbar\to\Lb\Lc K^-$ in the form of narrow $\Lc K^-$ resonances at 3800 and 3950~MeV.  Public LHCb data are used here to compare three hypotheses with an unbinned, background-aware two-dimensional Dalitz likelihood.  The baseline contains nonresonant three-body phase space and an $N(1900)$ contribution in $\Lb K^-$.  A second model tests the two a-priori prediction masses directly, fixing $\Lc K^-$ resonances at 3800 and 3950~MeV while fitting their widths.  A third model keeps the 3800-MeV mass fixed, allows the second $\Lc K^-$ mass and width to float, and adds two exotic $\Lc\Lb$ contributions.  The second model gives a nominal Wilks-equivalent improvement of $4.02\sigma$ over the baseline, while the third gives a nominal value of $6.17\sigma$. The quoted significances are model-comparison diagnostics only: the resonance fractions are constrained to be non-negative,
some fitted widths reach parameter boundaries, the floating mass introduces a trials effect, and efficiency and systematic uncertainties are not yet incorporated.
\end{abstract}

\section{Motivation}
The publication of collider data provides a direct way for theoretical predictions to be tested outside experimental collaborations.  Public data do not replace a collaboration-level analysis, but they can indicate whether a proposed interpretation is incompatible with the data or remains sufficiently interesting to motivate a more complete study.  The LHCb Open Data Ntupling Service provides user-configured ROOT ntuples without requiring the complete experiment-specific software stack~\cite{NtuplingService,CERNOpenData}.

The present analysis was initiated to test a prediction made at HADRON 2025 \cite{HADRON2025} and later presented to LHCb~\cite{LHCbTalk}.  The prediction concerned
\begin{equation}
  \Bbar\to\Lb\Lc K^-.
\end{equation}
BaBar had previously observed this decay and reported an enhancement in the $\Lc K^-$ invariant-mass distribution above approximately 3.5~GeV~\cite{BaBar}, stating ``A possible explanation for this [enhancement] is a resonant decay via a baryon resonance that has not yet been observed.''  At HADRON 2025, it was proposed that the enhancement was due to two baryon resonances rather than one.  The prediction was: ``More statistics for $\Bbar\to\Lb\Lc K^-$ will reveal narrow 5-sigma resonances in $\Lc K^-$ at 3.8 and 3.95~GeV'' \cite{HADRON2025}.

The analysis is organized as a hierarchy of three models.  The first asks how well a minimal nonresonant-plus-$N(1900)$ description accounts for the selected sample.  The second tests the two a-priori $f$-baryon mass hypotheses.  The third asks whether allowing the 3950 mass to vary and including two exotic $\Lc\Lb$ resonances can generate a better fit to the data.

The two $\Lc\Lb$ resonances in the third model are motivated by $f$-quark interpretations of experimental observations.  In the decay $B^-\to\Lc\bar p\pi^-$, Belle reported a pronounced low-mass enhancement in $\Lc\bar p$ that could be parameterized by a Breit--Wigner (BW) resonance with a mass of $3.35^{+0.01}_{-0.02}\pm0.02$~GeV and a width of $0.07^{+0.04}_{-0.03}\pm0.04$~GeV~\cite{Belle3350}.  Both BaBar and LHCb observed a similar low-mass enhancement, but they treated it as a threshold enhancement rather than a BW resonance \cite{BaBar3350,LHCb3350}. Within the $f$-quark model of \cite{new-quark}, a related state is expected in the $\Lambda_c^+\bar\Lambda$ channel.  Because 3350~MeV lies below the physical $\Lc\Lb$ threshold, only its above-threshold tail enters this decay.  

The second $\Lc\Lb$ resonance is $T_{c\bar c}(4020)^+$, for which Particle Data Group quotes a mass near 4024.1~MeV and width near 13~MeV~\cite{PDG2026,PDG4020}.  Through mechanisms described in \cite{new-quark}, it is proposed that this hadron also has the possibility of decaying to $\Lc\Lb$.

The objective of this paper remains deliberately limited: the analysis is a model-comparison study using public data, not a claim of evidence or observation of a new particle.

\section{LHCb open-data sample and offline selection}
A request to the LHCb Ntupling Service was configured for
\begin{equation}
B^0\to (\Lc\to pK^-\pi^+)\,(\Lb\to\bar p\pi^+)\,K^-,
\end{equation}
with charge-conjugate candidates included by the underlying LHCb selection \cite{Ntuple44}.  The original request returned more than 19 million candidates from 2011, 2012, and 2015--2017 data.  The relevant public stripping configuration includes, among other requirements, a $B$-candidate vertex-quality restriction and $\chi^2_{\rm IP}(B)<25$~\cite{StrippingLine}.  Particle-identification (PID) quantities are based on the standard LHCb calibration and PID machinery~\cite{PIDcalib}.

After downloading the ROOT files, the following cuts were made to the data:
\begin{align}
  {\rm ProbNN}_p &>0.50, &
  {\rm ProbNN}_K &>0.20, &
  {\rm ProbNN}_\pi &>0.30, \\
  \chi^2_{\rm vtx}(B^0)/{\rm ndf} &<10, &
  \chi^2_{\rm vtx}(\Lc)/{\rm ndf} &<8, &
  \chi^2_{\rm vtx}(\Lb)/{\rm ndf} &<8, \\
  \chi^2_{\rm FD}(B^0) &>20, &
  \chi^2_{\rm IP}(B^0) &<25.&&
\end{align}
where the first line are PID probability minima, the second are vertex quality cuts, and the last line are flight distance and impact parameter cuts.  15,168 events survived these cuts.  After these PID + vertex/displacement cuts, the following invariant-mass selections were made:
\begin{align}
 |M(\Lb)-m_{\Lambda}| &<5~\MeV, \\
 |M(\Lc)-m_{\Lc}| &<15~\MeV, \\
 |M(B^0)-m_{B^0}| &<30~\MeV.
\end{align}

\begin{figure}[H]
\centering
\includegraphics[width=0.88\linewidth]{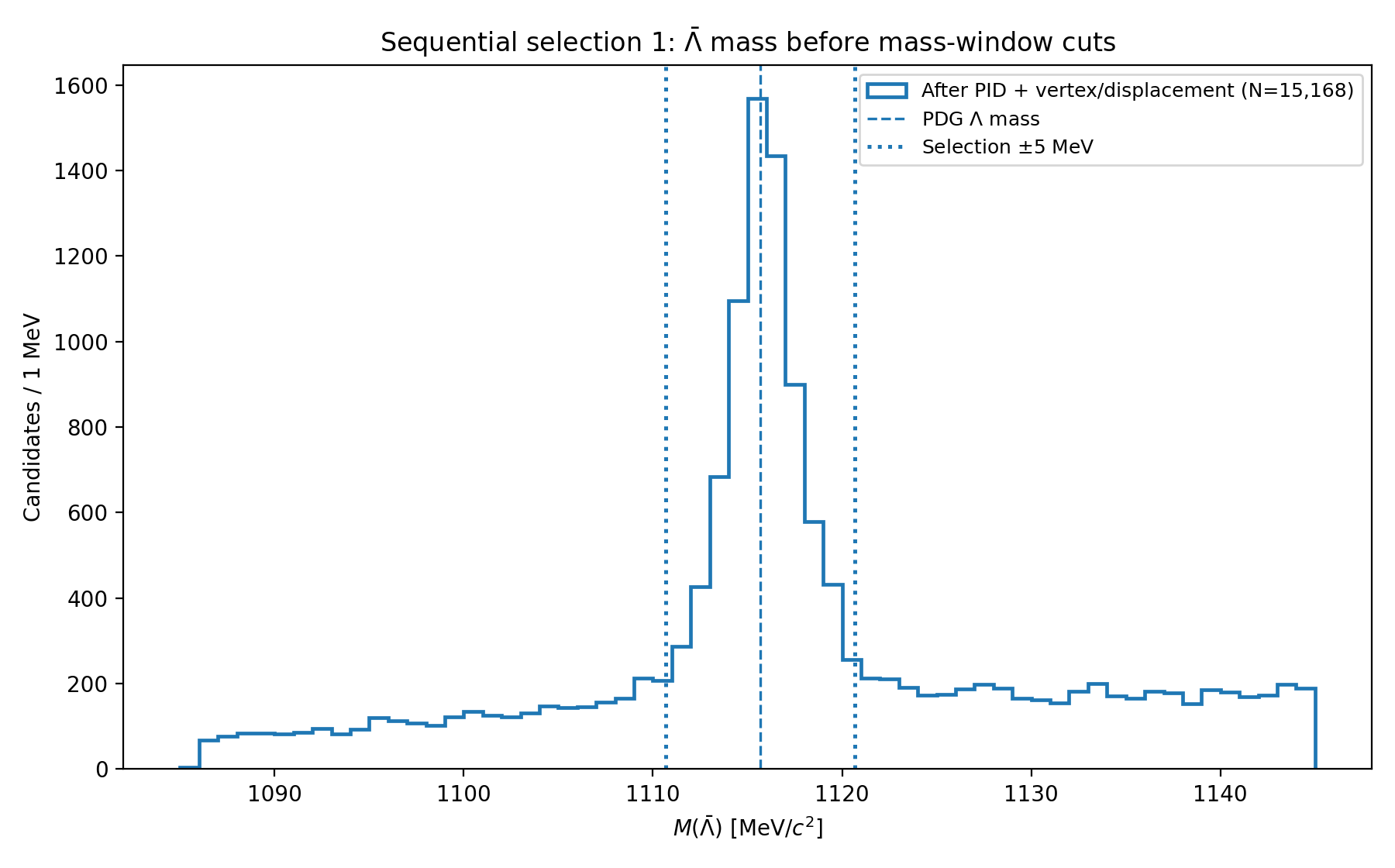}
\caption{$\Lb$ candidate mass before any invariant-mass window is applied.  The reduced sample already contains the PID and vertex/displacement requirements.  The dotted lines show the retained $\pm5$~MeV interval.}
\label{fig:selLambda}
\end{figure}

\begin{figure}[H]
\centering
\includegraphics[width=0.88\linewidth]{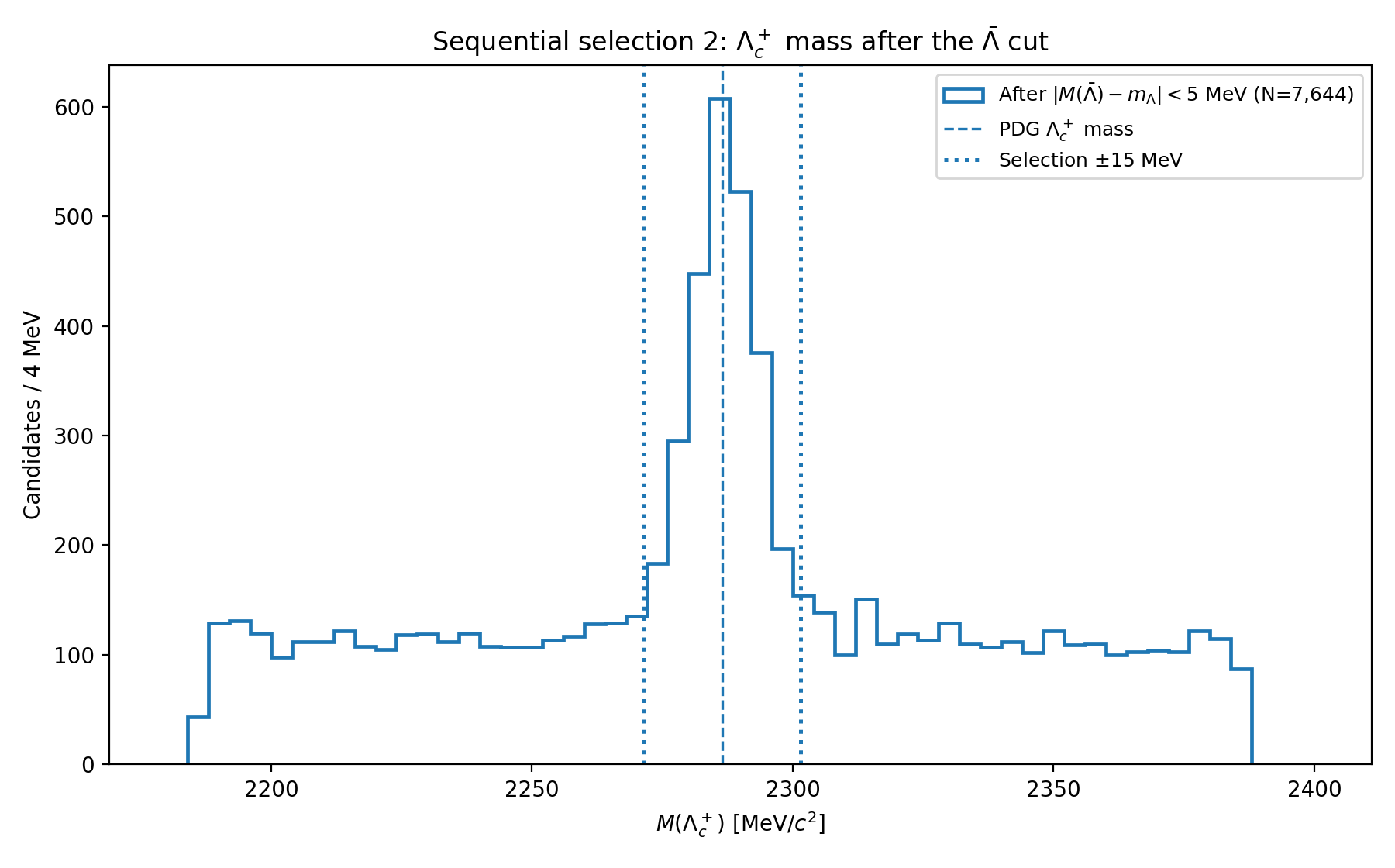}
\caption{$\Lc$ mass after the $\Lb$ mass requirement but before the $\Lc$ mass cut.  The dotted lines show the retained $\pm15$~MeV interval.}
\label{fig:selLc}
\end{figure}

\begin{figure}[H]
\centering
\includegraphics[width=0.92\linewidth]{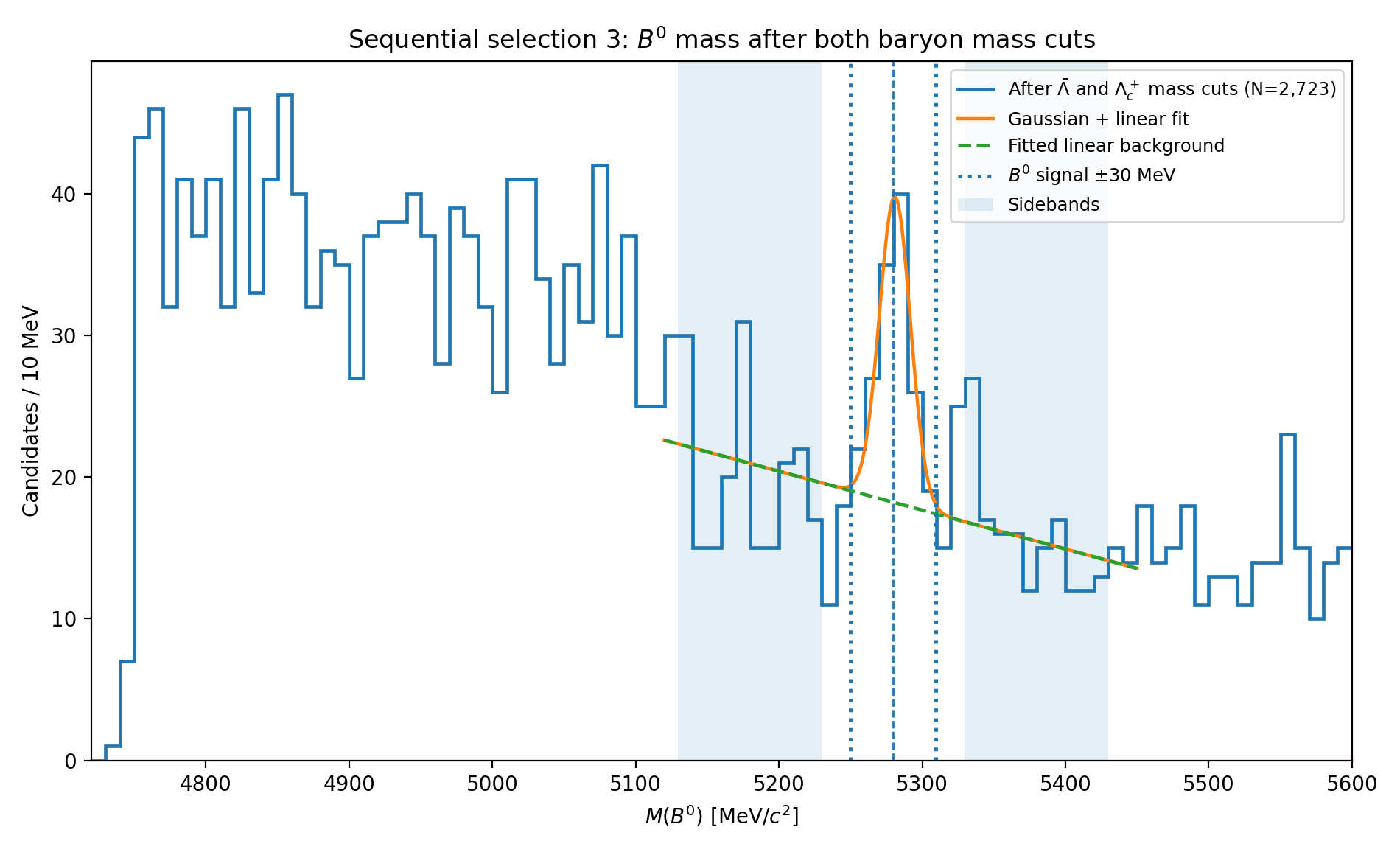}
\caption{$B^0$ mass after both baryon mass requirements but before the $B^0$ mass cut.  The Gaussian-plus-linear curve illustrates the signal peak on a falling combinatorial background.  The dotted lines delimit the $\pm30$~MeV signal region; the shaded intervals are the sidebands used to construct the Dalitz-background density.}
\label{fig:selB}
\end{figure}

Figures~\ref{fig:selLambda}--\ref{fig:selB} show the mass selections in the order in which they are applied, beginning with the $\Lb$ mass distribution.  Applying the $\pm5$~MeV $\Lb$ window leaves 7644 candidates shown in the $\Lc$ invariant mass plot.  Applying the $\pm15$~MeV $\Lc$ window leaves 2723 candidates shown in the $B^0$ plot.  Applying the $\pm30$~MeV $B^0$ window leaves 170 candidates in the signal region and 361 candidates in the sideband regions.

\section{Unbinned two-dimensional Dalitz likelihood}
For a three-body decay the three pairwise invariant masses are constrained by
\begin{equation}
 M^2(\Lc\Lb)+M^2(\Lc K^-)+M^2(\Lb K^-)
 =m_{B^0}^2+m_{\Lc}^2+m_{\Lambda}^2+m_K^2.
\end{equation}
Only two invariant masses are therefore independent.  The likelihood is evaluated event-by-event in the two-dimensional Dalitz coordinates
\begin{equation}
 \Phi=\bigl(M(\Lb K^-),\,M(\Lc K^-)\bigr),
\end{equation}
while $M(\Lc\Lb)$ is fixed by the kinematic relation for each event.  The histograms shown later are projections only; the fitted parameters are determined by the unbinned likelihood.

The likelihood for the 170 $B^0$-window candidates is
\begin{equation}
 -\ln\mathcal L=-\sum_i \ln\left[(1-f_B)S(\Phi_i)+f_B B(\Phi_i)\right].
\end{equation}
As illustrated in Fig.~\ref{fig:selB}, the $B^0$ distribution after both daughter-baryon mass cuts contains a narrow peak on a background that falls across the signal region.  The Gaussian-plus-linear description is used to quantify that sloping component.  The resulting fixed combinatorial-background fraction is
\begin{equation}
f_B=0.6478
\end{equation}
in the $\pm30$~MeV signal window.  The two-dimensional background density $B(\Phi)$ is obtained from the $B^0$ sidebands.  Because only 361 sideband candidates remain after the daughter-baryon mass requirements, a lightly smoothed histogram retains visible Poisson-scale lumps.  The nominal background model therefore applies a two-dimensional Gaussian smoothing kernel with $\sigma=3$ histogram bins, which suppresses these local fluctuations while preserving the broad Dalitz gradients.  Fits with weaker smoothing were also examined; the principal structures persist, while the weaker smoothing transfers visible sideband statistical fluctuations into the background density $B(\Phi)$.

Signal normalization is obtained numerically with phase-space Monte Carlo events distributed over the physical Dalitz region.  The resonance line shapes are convolved with a Gaussian mass resolution of
\begin{equation}
 \sigma_m=3~\MeV.
\end{equation}
For an above-threshold two-body resonance the simplified P-wave line shape used in this exploratory fit is
\begin{equation}
 BW_1(m)=\frac{M_R\Gamma_R}{M_R^2-m^2-iM_R\Gamma(m)}\left(\frac{q}{q_0}\right),
\end{equation}
with
\begin{equation}
 \Gamma(m)=\Gamma_R\left(\frac{q}{q_0}\right)^3\frac{M_R}{m}.
\end{equation}
The 3350-MeV pole is below $\Lc\Lb$ threshold, so the $q/q_0$ form cannot be normalized at its pole.  Its contribution is therefore represented only by the physical above-threshold tail of a constant-width relativistic Breit--Wigner,
\begin{equation}
 BW_{3350}(m)=\frac{M_X\Gamma_X}{M_X^2-m^2-iM_X\Gamma_X},
 \qquad M_X=3350~\MeV,\quad \Gamma_X=70~\MeV.
\end{equation}
All components in the present three-model comparison are added as noninterfering intensity terms.  This restriction is intentional: it keeps the parameter counting transparent and avoids introducing unconstrained phases into the current low-statistics sample.

\section{Three nested model tests}
The baseline hypothesis is
\begin{equation}
 H_0:\qquad S_0={\rm NR}+N(1900),
\end{equation}
where NR denotes three-body phase space and $N(1900)$ is a $\Lb K^-$ contribution with fixed nominal mass and width $M=1911$~MeV and $\Gamma=292$~MeV in the implementation used here.

The second hypothesis tests the two masses specified before examination of the present open-data sample,
\begin{equation}
 H_1:\qquad S_1=S_0+R_{3800}+R_{3950},
\end{equation}
with $M_1=3800$~MeV and $M_2=3950$~MeV fixed and both widths fitted.  These are the masses stated in the HADRON 2025 prediction.

The third hypothesis allows the mass of the second $\Lc K^-$ resonance to vary between 3850--4050~MeV and adds the two $\Lc\Lb$ resonances mentioned above:
\begin{equation}
 H_2:\qquad S_2=S_0+R_{3800}+R_2(M_2,\Gamma_2)
 +X_{3350}+T_{c\bar c}(4020)^+.
\end{equation}
The 3800-MeV mass remains fixed, and the 3350/70 pole and $T_{c\bar c}(4020)^+$ are fixed to the values quoted above.

The masses, widths and likelihoods of the fits are shown in Table~\ref{tab:parameters}, while the mixture fractions are shown in Table~\ref{tab:fractions}.  They are model fractions within the fitted signal probability density function (PDF) and should not be identified directly with branching fractions.

\begin{table}[h]
\centering
\caption{Best-fit parameters for the three models.  Masses and widths are in MeV. ``Boundary'' marks a parameter at an imposed fit limit.}
\label{tab:parameters}
\begin{tabular}{@{}lccc@{}}
\toprule
Parameter & $H_0$ & $H_1$ & $H_2$ \\
\midrule
$M_{3800}$ & -- & 3800 fixed & 3800 fixed \\
$\Gamma_{3800}$ & -- & 0.30 (boundary) & 0.30 (boundary) \\
$M_2$ & -- & 3950 fixed & $3996.2$ \\
$\Gamma_2$ & -- & 100 (boundary) & 19.75 \\
$M_{3350}$, $\Gamma_{3350}$ & -- & -- & 3350, 70 fixed \\
$M_{4020}$, $\Gamma_{4020}$ & -- & -- & 4024.1, 13 fixed \\
$-\ln\mathcal L$ & 2420.094 & 2406.988 & 2390.990 \\
\bottomrule
\end{tabular}
\end{table}

\begin{table}[h]
\centering
\caption{Best-fit signal-model fractions.}
\label{tab:fractions}
\begin{tabular}{@{}lrrr@{}}
\toprule
Component & $H_0$ & $H_1$ & $H_2$ \\
\midrule
NR & 77.19\% & 32.32\% & 0.20\% \\
$N(1900)$ & 22.81\% & 24.52\% & 29.47\% \\
$R_{3800}$ & -- & 5.18\% & 4.23\% \\
$R_{3950}$ or floating $R_2$ & -- & 37.98\% & 10.83\% \\
3350/70 tail & -- & -- & 52.07\% \\
$T_{c\bar c}(4020)^+$ & -- & -- & 3.19\% \\
\bottomrule
\end{tabular}
\end{table}

In both the second and third models, the fitted width of the 3800~MeV resonance ends up below the assumed detector resolution and is therefore treated as unresolved. Contrastingly, in $H_1$ the width of the 3950-MeV resonance reaches the 100-MeV upper limit imposed by the fit.  This indicates that, in the restricted two-resonance model, the 3950 term is being used as a broad shape rather than determining a well-localized resonance. In $H_2$, after allowing the mass to move up from 3950 to 3996~MeV (and including the additional exotic $\Lc\Lb$ resonances), the width narrows to 19.8~MeV.

\section{Likelihood comparisons}
The direct likelihood-ratio comparisons are
\begin{align}
 2\Delta\ln\mathcal L(H_1/H_0)&=26.21,\\
 2\Delta\ln\mathcal L(H_2/H_0)&=58.21,\\
 2\Delta\ln\mathcal L(H_2/H_1)&=32.00.
\end{align}
With nominal parameter counting, $H_1$ has four additional free parameters relative to $H_0$, while $H_2$ has seven additional parameters relative to $H_0$.  The standard Wilks assumptions are not
strictly satisfied because the resonance fractions are constrained to
be non-negative and the resonance widths are not identifiable under
the null hypothesis. Nevertheless, as a diagnostic of the relative
fit improvements, a naive Wilks conversion gives
\begin{align}
 H_1/H_0:&\qquad p_{\rm naive}=2.87\times10^{-5},\qquad Z_{\rm naive}=4.02\sigma,\\
 H_2/H_0:&\qquad p_{\rm naive}=3.44\times10^{-10},\qquad Z_{\rm naive}=6.17\sigma.
\end{align}
For reference, the nominal $H_2/H_1$ conversion is $Z_{\rm naive}=4.88\sigma$ for three additional parameters.

These Gaussian-equivalent values are not calibrated experimental significances.  Model choices were made about which exotic hadrons to include, and the floating $M_2$ in $H_2$ introduces a look-elsewhere effect that has not been taken into account. 

\section{Two-body projections}
\begin{figure}[H]
\centering
\includegraphics[width=0.96\linewidth]{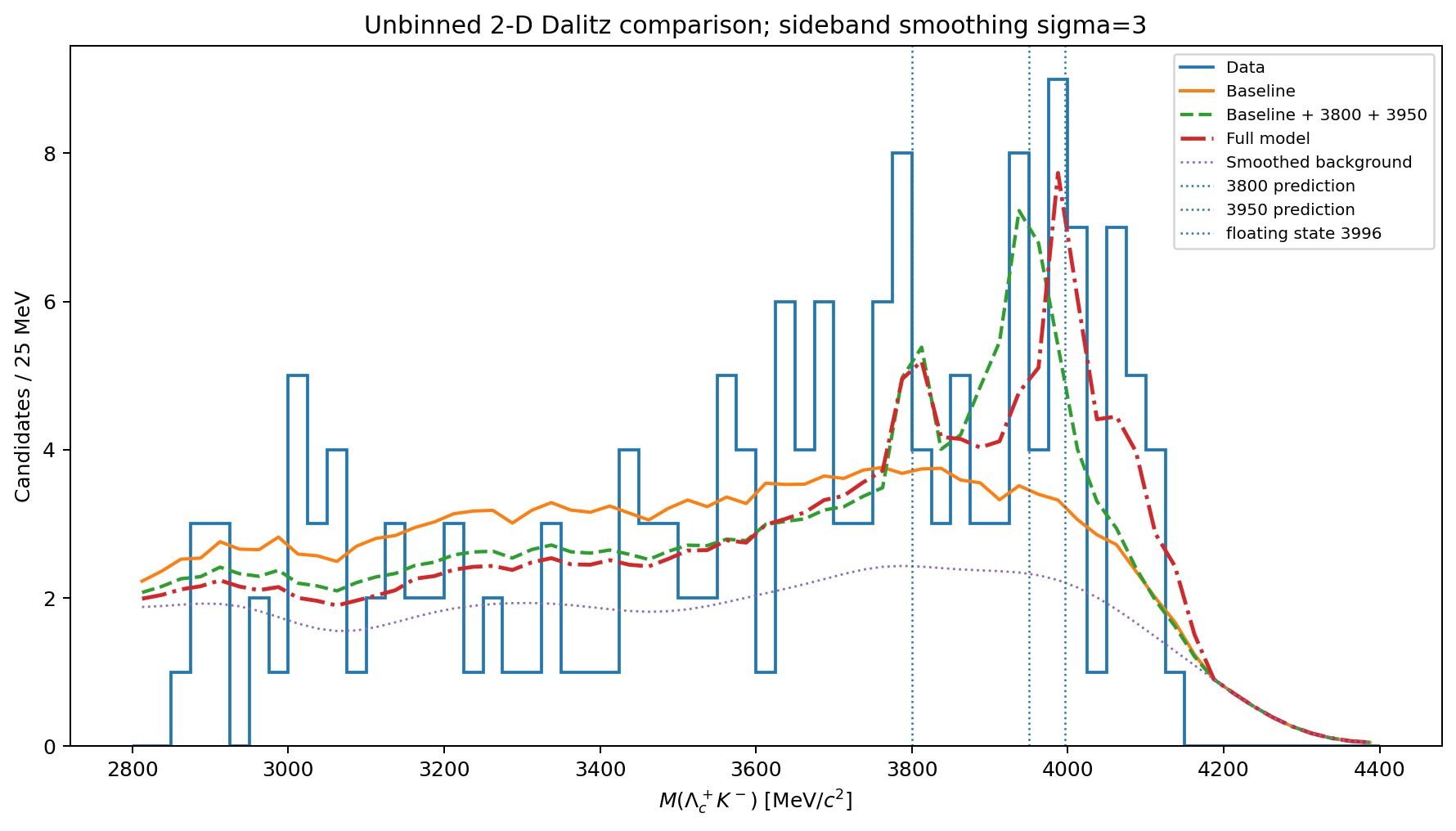}
\caption{$\Lc K^-$ projection for the baseline, the fixed 3800+3950 prediction model, and the full model in which the second mass floats and the 3350 tail and $T_{c\bar c}(4020)^+$ are included.  The background density uses the nominal $\sigma=3$ two-dimensional sideband smoothing.}
\label{fig:LcK}
\end{figure}

\begin{figure}[H]
\centering
\includegraphics[width=0.96\linewidth]{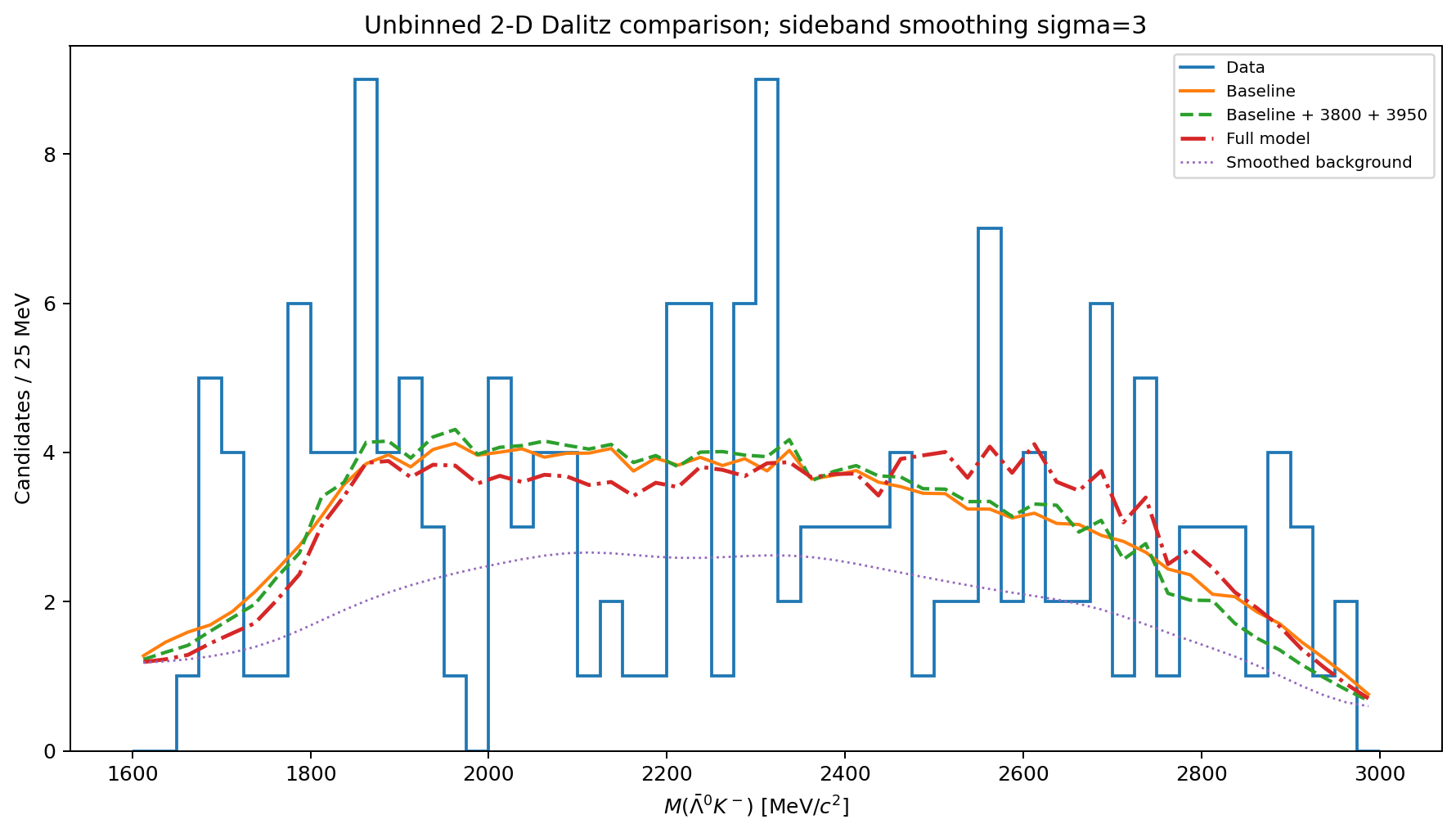}
\caption{$\Lb K^-$ projection.  Changes caused by $\Lc K^-$ and $\Lc\Lb$ components appear through the correlated three-body kinematics.}
\label{fig:LamK}
\end{figure}

\begin{figure}[H]
\centering
\includegraphics[width=0.96\linewidth]{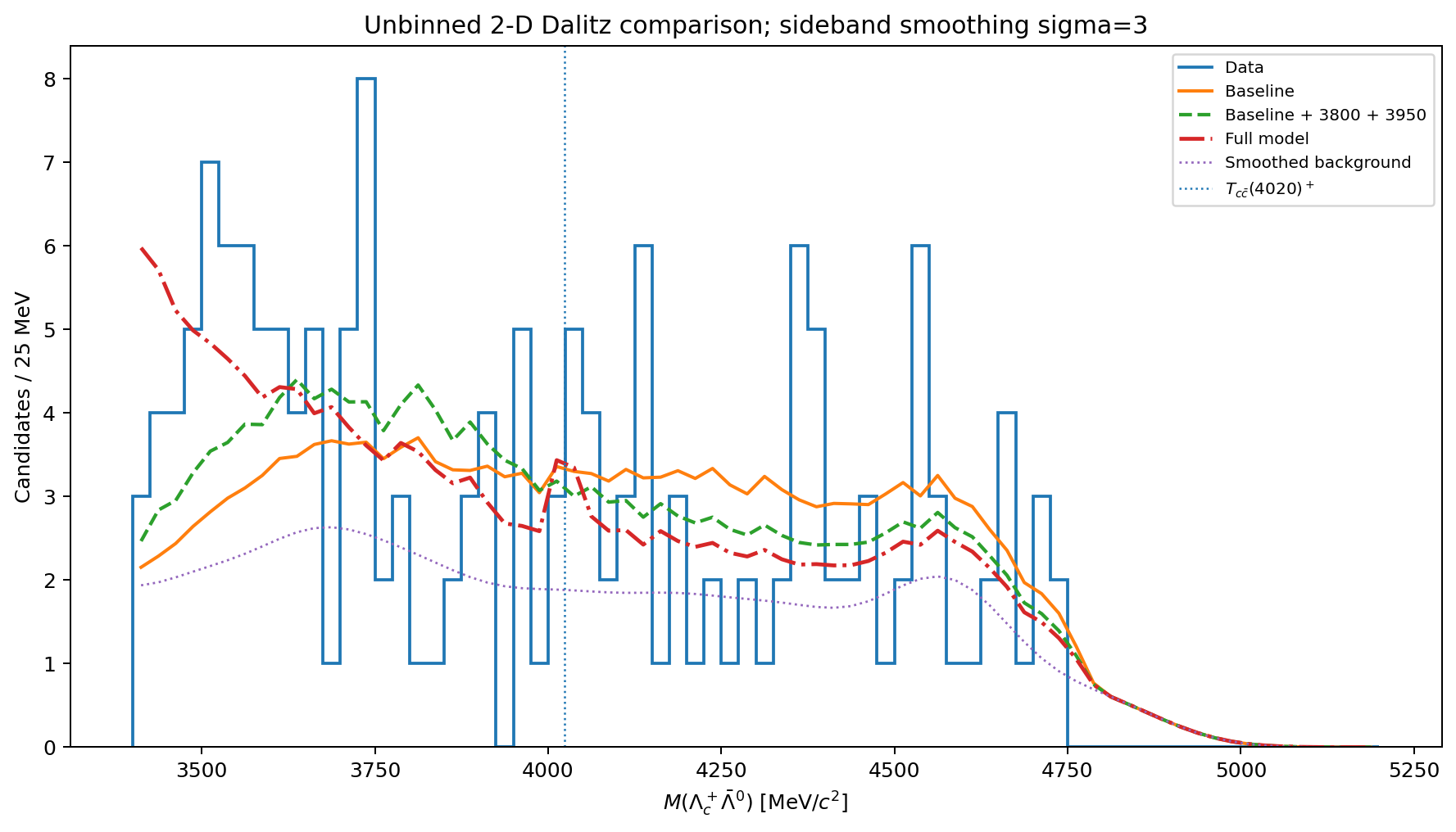}
\caption{$\Lc\Lb$ projection.  The full model contains the physical tail of the fixed 3350/70-MeV subthreshold pole and a fixed $T_{c\bar c}(4020)^+$ contribution.}
\label{fig:LcLam}
\end{figure}

Figures~\ref{fig:LcK}--\ref{fig:LcLam} show projections of the three fitted two-dimensional models.  The curves are not independent one-dimensional fits; each is obtained by projecting the corresponding Dalitz PDF after the unbinned fit.

\section{Discussion}
The three models considered address two different questions.  First, the pre-specified 3800- and 3950-MeV $\Lc K^-$ hypotheses improve the baseline likelihood by $2\Delta\ln\mathcal L=26.21$.  These are the masses stated in the HADRON 2025 prediction and neither was selected by scanning the present sample.  The 3950 component is nevertheless very broad in this restricted model, reaching the imposed 100-MeV width ceiling, while the 3800 component is unresolved.  The result therefore supports additional $\Lc K^-$ structure more clearly than it measures a resonance width at exactly 3950~MeV.

Second, the full model changes the interpretation of that broad component.  A fixed 3350/70-MeV subthreshold tail accounts for roughly half of the fitted signal density, while $T_{c\bar c}(4020)^+$ contributes a few percent.  Once those correlated $\Lc\Lb$ structures are present, the second $\Lc K^-$ term no longer needs to be broad at 3950~MeV.  Instead it moves to $3996.2$~MeV with a fitted width of about $19.8$~MeV.  The comparison therefore suggests that the location and width assigned to the second $\Lc K^-$ contribution are sensitive to the treatment of the crossed $\Lc\Lb$ channel.

The large fitted 3350-tail fraction should not by itself be read as evidence that the Belle enhancement and the present $\Lc\Lb$ distribution share a common pole.  The pole is below the physical $\Lc\Lb$ threshold, the current model uses only its phenomenological Breit--Wigner tail, and no coupled-channel or final-state-interaction calculation is included.  Its role here is to test a concrete alternative to an arbitrary smooth threshold function.  The fact that a fixed 3350/70 tail produces a large likelihood improvement makes that hypothesis worth examining with a more complete amplitude treatment, but does not establish the interpretation.

Several limitations remain.  The current likelihood does not contain an efficiency map over the Dalitz plane, does not use event-by-event mass uncertainties, and treats the principal signal components as noninterfering intensities.  The 3-MeV resolution is an approximate common convolution rather than a detector-calibrated resolution model for each candidate.  The sideband background density and its fraction are fixed rather than profiled simultaneously with the signal.  Finally, the selection itself was optimized using control distributions, so a complete frequentist calibration should reproduce the analysis procedure in pseudoexperiments where relevant.

For these reasons the strongest defensible conclusion is comparative: the selected public LHCb sample is described substantially better by the extended hypotheses with two $\Lc K^-$ structures than by the chosen baseline.  Also, the data favor the second $\Lc K^-$ mass being nearer 4.00~GeV once the proposed exotic $\Lc\Lb$ structures are included.  A collaboration-level amplitude analysis with substantially larger statistics is required to determine whether any of these structures correspond to genuine hadronic resonances.

\section{Conclusion}
A prediction made at HADRON 2025 of $f$-baryon resonances near 3.8 and 3.95~GeV in $\Lc K^-$ has been tested using public LHCb data for $\Bbar\to\Lb\Lc K^-$.  Relative to the NR+$N(1900)$ baseline, the prediction was tested by adding fixed 3800- and 3950-MeV $\Lc K^-$ states.  This addition generated $2\Delta\ln\mathcal L=26.21$ and a nominal diagnostic significance of $4.02\sigma$.  A third exploratory extension model added a fixed 3350/70-MeV subthreshold $\Lc\Lb$ tail and a $T_{c\bar c}(4020)^+$ resonance, while allowing the second $\Lc K^-$ mass to float.  It generated $2\Delta\ln\mathcal L=58.21$ relative to baseline and a nominal diagnostic significance of $6.17\sigma$.  In the third model, the floating state moved to $3996.2$~MeV with a fitted width of $19.8$~MeV, while the width of the 3800-MeV contribution was too narrow to be resolved at the assumed 3-MeV resolution scale.  These likelihood improvements provide quantitative motivation for a higher-statistics, efficiency-corrected amplitude analysis, but do not constitute an experimental claim of evidence or observation.

\section*{Acknowledgments}
This research used open data provided by the LHCb Collaboration through the CERN Open Data Portal and the LHCb Open Data Ntupling Service.  The LHCb Collaboration and CERN are thanked for preserving and making these data publicly available.  This work is an independent analysis of publicly released LHCb data and is not an official result of the LHCb Collaboration.  Neither CERN nor the LHCb Collaboration has reviewed or endorsed the analysis or its conclusions.

\end{document}